\documentclass[12pt]{article} 
\usepackage{url}
\usepackage{hyperref}
\usepackage{fancybox}
\usepackage{graphicx}
\usepackage{wrapft}
\usepackage{caption}
\usepackage{subcaption}
\usepackage{epsf}
\usepackage{amsmath,amssymb}
\usepackage{mathrsfs}
\usepackage{comment,cite}

\DeclareMathOperator{\sgn}{sgn}
\usepackage{color}
\definecolor{rosso}{cmyk}{0,1,1,0.60}
\definecolor{rossos}{cmyk}{0,1,1,0.55}
\definecolor{rossoc}{cmyk}{0,1,1,0.2}
\definecolor{blus}{cmyk}{1,1,0,0.3}
\definecolor{blum}{cmyk}{1,0.76,0,0.2}
\definecolor{blu}{cmyk}{1.0,0.76,0,0}
\definecolor{bluc}{cmyk}{1,1,0,0.1}
\definecolor{verde}{cmyk}{0.92,0,0.59,0.25}
\definecolor{verdec}{cmyk}{0.92,0,0.59,0.15}
\definecolor{verdes}{cmyk}{0.92,0,0.59,0.4}
\definecolor{giallo}{cmyk}{0,0,1,0}
\definecolor{gialloverde}{cmyk}{0.44,0,0.74,0}
\begin{document} 
 
\mbox{ } \\[-1cm]
\mbox{ }\hfill TU-1185
\bigskip

\thispagestyle{empty}
\setcounter{page}{0}

\vspace{10mm}
\begin{center}
   {\Large{\sf  Long-lived Higgsino pairs decaying within the LHC detectors}}\\[10mm]
{\large 
Francesca~Borzumati$^{1}$, Kaoru~Hagiwara$^{2}$, Kentarou Mawatari$^3$,}\\[1.010ex] 
{\large
  Youichi~Yamada$^{4}$,}  and
{\large
  Toshifumi~Yamashita$^{5}$} 
\end{center}
\bigskip
\begin{center}
$^1$ {\it Shibaura Institute of Technology, Omiya Campus, Saitama 337-8570, Japan}\\[1mm] 
$^2${\it {\small KEK} Theory Division, Tsukuba 305-0801, Japan} \\[1mm] 
$^3${\it Faculty of Education, Iwate University, Morioka, Iwate 020-8550, Japan}\\[1mm] 
$^4${\it Department of Physics, Tohoku University, Sendai 980-8578, Japan} \\[1mm] 
$^5${\it Department of Physics, Aichi Medical University, Nagakute, Aichi 480-1195, Japan}
\end{center}
\vskip 2cm


\begin{abstract}
Pair-produced long-lived lightest Higgsinos decaying within a few meters in
 the \small{LHC} detectors are considered.
The relatively light nearly-pure Higgsino states object of this study
 are those of a supersymmetric minimal model with gauge-mediated
 supersymmetry breaking and heavy sfermions and gauginos.
The heavier Higgsinos decay instantly into the lightest one, 
 with very little activities.
Hence, all pair-produced Higgsinos end up as a pair of the lightest neutral
 Higgsinos, possibly long lived and decaying into the gravitino and either
 the Higgs or the Z boson.
If they decay inside the detectors, these bosons deposit huge energies in the
 calorimeters.
The cross section  $\sigma_2$ ($\sigma_1$) for the events with two (at least one) 
 decaying-in-flight lightest Higgsinos is examined. 
The combination $(\sigma_1)^2/\sigma_2$ is found to be almost insensitive to
 the lightest Higgsino lifetime, thereby providing a good variable to measure
 its mass.
The ratio $\sigma_1/\sigma_2$, instead, mildly dependent on its
 mass, can give good information on its lifetime. 
It is argued that the lightest Higgsinos in the events are rather slow,  as
 the very relativistic ones escape away.
Thus, the initial state radiations tend to be soft and a trigger system
 appropriate for such events is necessary to detect them.
\end{abstract}

\newpage
\setcounter{page}{1}
\setlength{\parskip}{1.01ex}

{\noindent \bf Introduction.}
In the last two decades weak-scale
 supersymmetry~(\small{SUSY})~\cite{Martin:1997ns} has been struggling to
 maintain consistency with experimental data, as the superpartners of the
 Standard Model~(\small{SM}) particles seem to be considerably heavier than
 the electroweak~(\small{EW}) scale.
Superficially, weak-scale \small{SUSY} with minimal particle
 content~({\small MSSM})~\cite{MSSMWorkingGroup:1998fiq}
 appears incapable of guaranteeing a natural solution to the little hierarchy
 problem~\cite{Barbieri:1987fn}.
Thus, \small{SUSY} model building has found itself in need to accommodate at
least a moderate amount of decoupling with respect to the~\small{EW} scale
of all, or of a subset of the~\small{SUSY} massive parameters.
See for example
  Refs.~\cite{Arkani-Hamed:2004ymt,Giudice:2004tc,Delgado:2007rz,Allanach:2016pam,Fukuda:2015pra}. 
Among these, noteworthy is Ref.~\cite{Fukuda:2015pra}, with a class of models 
 that predict heavy superpartners in a framework of gauge-mediated SUSY
 breaking~({\small GMSB}), while simultaneously providing some suppression
 of the parameter expressing the ensuing tuning.

Experimental constraints, however, seem to leave still open
 the possibility of a not too heavy parameter $\mu$ in the superpotential.
Thus, the Higgsino sector may be lighter than other sectors,
 hopefully providing a handle to track down these models. 
Many phenomenological studies have appeared in recent years on strategies to
 search for charginos and neutralinos with substantial Higgsino
 components, in scenarios in which  the lightest neutralino is
 stable~\cite{Kribs:2013lua,Han:2014kaa,Mahbubani:2017gjh,Fukuda:2019kbp,Dutta:2019gox,ATLAS:2019lng},
 or long-lived~\cite{Meade:2010ji,Liu:2015bma,Liu:2018wte}.
Following this trend, we concentrate here on extracting information
 relevant for the search and identification of long-lived lightest neutralinos at
 the {\small LHC}.

In what follows we define our framework.
Although inspired by the models in Ref.~\cite{Fukuda:2015pra}, this is here
 characterized as a simplified model~\cite{Alwall:2008ag}, which may 
 encompass the description of the lower-energy physics of existing and also
 future underlying \small{SUSY} models with possible
 new solutions to the naturalness problem.
We give cross sections and distributions for the pair production of the lightest
 neutralino, investigate its long-lived decays and search for variables 
 giving informations on the parameters of our model.
We conclude with a summary of our work.
\\

{\noindent \bf Nearly-pure Higgsinos.}
Our simplified model is that of nearly-pure Higgsinos.
We assume an {\small{MSSM}} with {\small GMSB}, which predicts
 the gravitino~$\widetilde{G}$ as the lightest
 supersymmetric particle~({\small LSP}).
We are noncommittal about the mechanisms that allow the splittings/decouplings
 of sfermions and gauginos, but we assume that heavy top-squark masses,
 together with a large value of $\tan \beta$, reproduce
 the correct value for the mass of the {\small{SM}}-like Higgs, $h$, whereas the other
 Higgs states $H$, $A$, and $H^\pm$ remain heavy.
(This is achieved in  Ref.~\cite{Fukuda:2015pra} by using non-minimal {\small GMSB}
  messengers~\cite{Buican:2008ws}, which in turn trigger a focus point
  mechanism~\cite{Feng:1999mn} in the evolution from the messenger scale to the
  {\small{EW}} scale.)
We also assume that the gauginos are sufficiently decoupled to obtain nearly-pure
 Higgsino states.
These are: two neutral states, $\widetilde{H}^{0}_{1,2}$, with
 $\widetilde{H}^{0}_{1}$ the lighter of the two, or
 next-to-the \small{LSP} (\small{NLSP}), and two charged ones,
 $\widetilde{H}^{\pm}$.
The states $\widetilde{H}^{0}_{1,2}$ are the neutralinos
 $\widetilde{\chi}^0_{1,2}$ obtained rotating  the gaugino-Higgsino basis
 $(\widetilde{B}, \widetilde{W}, \widetilde{H}^0_d, \widetilde{H}^0_u)$
 by the diagonalization matrix $N$~\cite{Martin:1997ns}: they are linear
 combinations of $\widetilde{H}^{0}_u$ and $\widetilde{H}^{0}_d$, with maximal
 mixing, duly multiplied by factors of $N_{1,i}$ and $N_{2,i}$ (with
 $N_{1,i},\,N_{2,i} \simeq 0$ for $i=1,2$). 
Similarly, the states $\widetilde{H}^{\pm}$ are the chargino mass eigenstate
 $\widetilde{\chi}^{\pm}_1$.
We use nevertheless the symbols $\widetilde{H}_1^{0}$, $\widetilde{H}_2^{0}$,
 and $\widetilde{H}^{\pm}$ as a reminder of the particular nature of the
 charginos and neutralinos object of this study.
These states are nearly degenerate, with masses in ascending order,
 $m_{\widetilde{H}^0_1}  <  m_{\widetilde{H}^{\pm}} < m_{\widetilde{H}^0_2}$,
 and mass splittings $\Delta m_0$ and $ \Delta m_{\pm}$ given, for
 $\sin\beta\sim\!1$, by~\cite{Fukuda:2019kbp}:
\begin{equation}
  \Delta m_0   \equiv  m_{\widetilde{H}_2^0}   -\! m_{\widetilde{H}_1^0} 
  \simeq
  m_Z^2  \left[ \frac{c^2_W}{M_2} + \frac{s^2_W}{M_1} 
    \right], 
  \quad \quad \,
  \Delta m_{\pm}  \equiv  m_{\widetilde{H}^{\pm}} -\! m_{\widetilde{H}_1^0}
  \simeq  \frac{1}{2} \Delta m_0 ,
\label{eq:splittHiggs}  
\end{equation}
 where $s_W$ and $c_W$ are the sine and cosine of the Weinberg angle,
 $M_2$, $M_1$ the masses of the {\small SU(2)}, {\small U(1)} gauginos.
These expressions for $\Delta m_0$ and $\Delta m_{\pm}$ are at the tree level and
 do not include higher-order corrections which are, however, used for the determination
 of the points of parameter space in Ref.~\cite{Fukuda:2015pra}.

The parameters in our analysis are then:
 $m_{\widetilde{H}^0_1}$ ($\simeq\vert\mu\vert$), $\Delta m_0$,
 and $\langle F \rangle$, the {\small{SUSY}}-breaking auxiliary
 {\it vacuum expectation value}~({\it v.e.v.})  that enters in the Higgsino-gravitino
 interaction, well approximated by the interactions between Higgsinos and the
 massless Goldstino $\widetilde{G}_0$~\cite{Cremmer:1982en}:
\begin{equation}
 {\cal{L}} \ \supset \ 
  \frac{1}{\langle F \rangle} (\partial_{\mu} \,\widetilde{G}_0)^{\alpha}
  \left[
    \left(\sigma^{\nu} \bar{\sigma}^{\mu} \widetilde{H}_j\right)_{\!\alpha}
    (D_{\nu} H_j^{\ast})
    + i \left(\sigma^{\mu} \widetilde{H}_j\right)_{\!\alpha}
    \frac{\delta W^{\ast}(H_u^\ast,H_d^\ast,\cdots)}{\delta (H_j^{\ast})}
  \right].    
\label{eq:LagGOLDSTINO}
\end{equation}
Here the index $j$ runs over the two Higgs doublet superfields $H_u$ and $H_d$ 
 denoted, as usual, by the same symbols assigned to their scalar components,
 while $\widetilde{H}_j$ are their fermionic components;
 $\widetilde{G}_0$ is the longitudinal part of  $\widetilde{G}$; 
 $W$, the superpotential; $D_{\nu}$,  the covariant derivative with respect to the
 gauge symmetries obeyed by the Higgs superfields, i.e. 
 $D_{\nu}= \partial_{\nu} +i g W^a_{\nu} T^a + ig^{\prime} Y B_{\nu}$;
 the various  $\sigma$ are Pauli matrices, and the indices $\alpha$, spinor
 indices.

Two additional parameters $\sgn{\mu}$ and $\tan\beta$ (likely to be large,
 as mentioned) enter typically in
 the factors of rotation from current to mass eigenstates.
We do not investigate them here, except in the case of the {\small{NLSP}}
 branching ratios, as will be seen later in this section. 
The other parameters of the underlying {\small{GMSB MSSM}} 
 determine the heavy masses of all other superpartners.
As mentioned, these are practically decoupled from our framework, which
 consists then in the \small{SM} enlarged by a gaugino-unmixed Higgsino sector
 and the gravitino.

We take $m_{\widetilde{H}^0_1}$ in the interval $[100\!-\!1000]\,$GeV.
The lower end of this interval is amply excluded by {\small{CMS}}~\cite{CMS:2021cox}
 and {\small{ATLAS}}~\cite{ATLAS:2022rcw}, in the case of promptly
 decaying {\small{NLSP}}s, while the {\small LHC} exclusion of long-lived
 {\small{NLSP}}s is a complicated function of their lifetime.

We assume the small splitting $\Delta m_0$ in Eq.~(\ref{eq:splittHiggs})
 to be large enough (it is a few GeV in the models of
 Ref.~\cite{Fukuda:2015pra}) to render prompt the {\small{EW}} decays:
 $ \widetilde{H}^{\pm} \to  \widetilde{H}^{0}_{1}\,W^{\pm\,\ast}\!$,
 $\ \widetilde{H}^{0}_{2} \to  \widetilde{H}^{0}_{1}\,Z^\ast\!$,
 $\ \widetilde{H}^{0}_{2} \to  \widetilde{H}^{\pm}\,W^{\mp\,\ast}\!$,
 with off-shell  $W$ and $Z$ bosons emitting pairs of quarks or leptons.
For values of $\Delta m_0 $, larger than the {\small{QCD}} scale, the second
 of these decays,
 $\widetilde{H}^{0}_{2} \to \widetilde{H}^{0}_{1}\,Z^\ast \to \widetilde{H}^{0}_1\, f \bar{f}$,
 for example, has width: 
 $  \sim  n_0 \times  ({G_F^2}/{15 \pi^3})\times (\Delta m_{0})^5 $
 with omitted factors of rotations from currents to mass eigenstates and 
 $n_0$ counting the number of $f\bar{f}$ pairs produced by the exchange of the
 $Z$ bosons, weighted by their color, charge and isospin. 
 (If the splittings $\Delta m_0$ and $\Delta m_{\pm}$ are very small 
 these decays are displaced.
 See Refs.~\cite{Mahbubani:2017gjh,Dutta:2019gox,Fukuda:2019kbp}.)

\begin{figure}[t!]
\begin{center}
   \begin{subfigure}[b]{0.41\textwidth}
    \includegraphics[height=6.75cm]{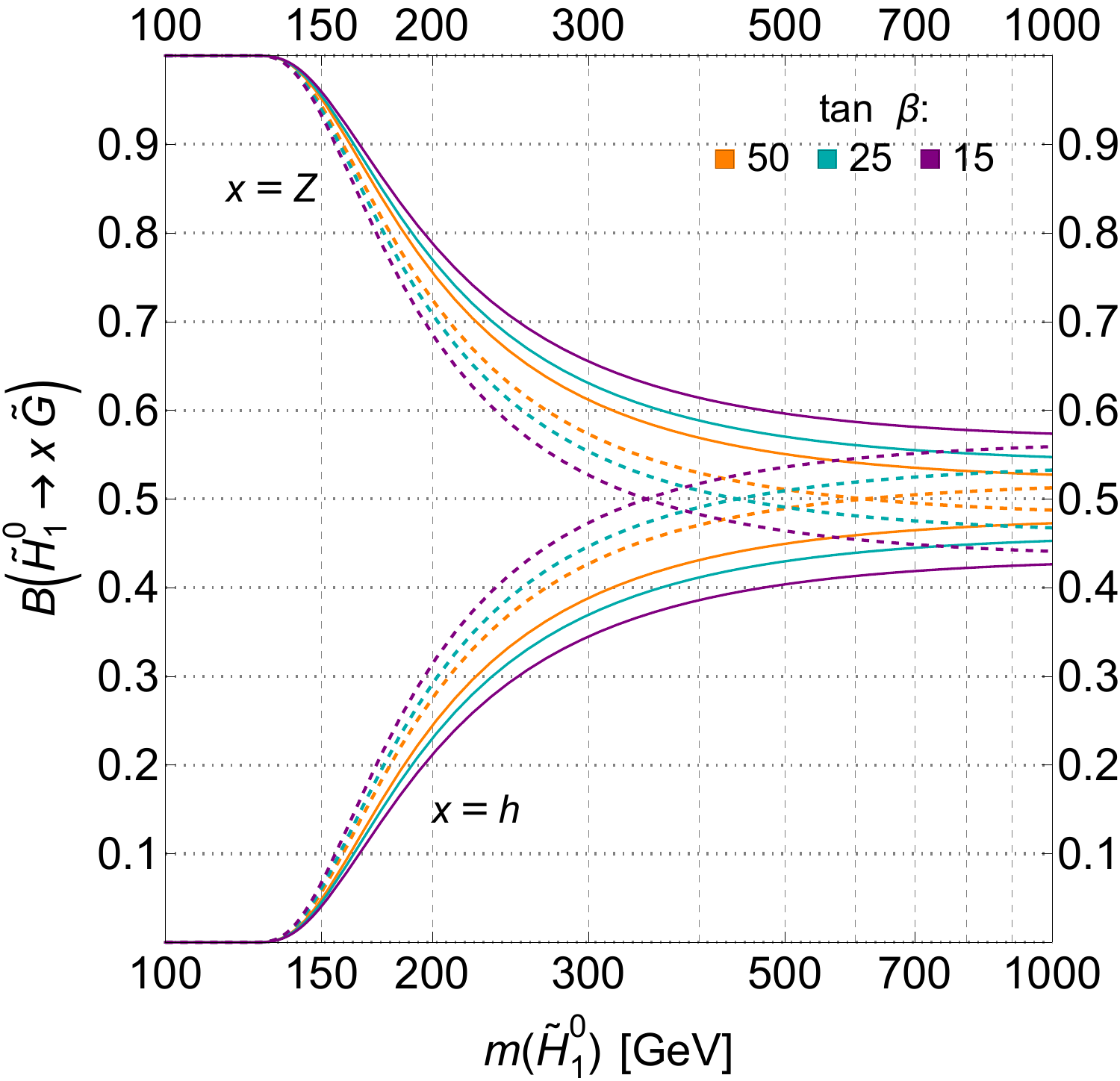}
    \caption{ } 
    \label{fig:f1a}
  \end{subfigure}
   \quad  \quad  \quad
  \begin{subfigure}[b]{0.41\textwidth}
    \includegraphics[height=6.75cm]{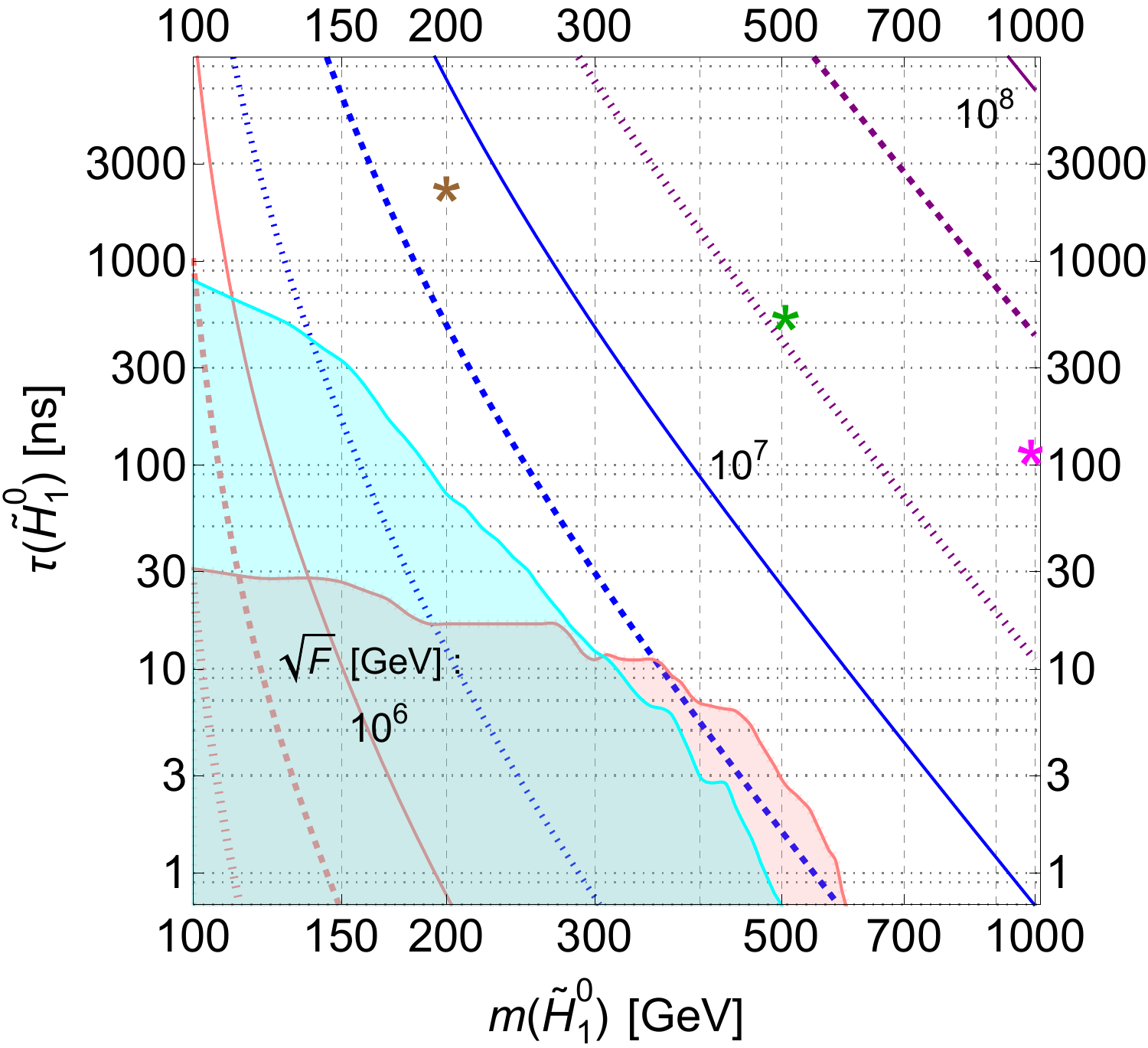}
    \caption{ }
    \label{fig:f1b}
  \end{subfigure}
\end{center}  
\caption{\small{(a) Dependence on  $\tan \beta$ and $\sgn{\mu}$
 of the branching ratios $B(\widetilde{H}^0_1\to x \widetilde{G})$.
Descending curves have $x= Z$, ascending ones, $x=h$.
In the same group, solid (dashed) curves with the same color correspond
 to positive (negative) $\mu$; with different colors to different values of
 $\tan \beta$.
 (b) Dependence on  $\langle F \rangle$ of the \small{NLSP} lifetime,
 for $\sin \beta \sim 1$, 
 shown for $\sqrt{\langle F \rangle} =10^8$, $10^7\,$,$10^6\,$GeV
 by the purple, blue, pink solid lines.
For each of these three curves, with $\sqrt{\langle F \rangle} =10^n\,$GeV,
 the dotted, dashed curves of the same color, have respectively
 $\sqrt{\langle F \rangle} = 2\times10^{(n-1)}$, $5\times 10^{(n-1)}\,$GeV.
 See text for details on the excluded cyan- and pink-shaded regions and
 on the three star points.
  }} 
\label{fig:BRLFT}
\end{figure}

The various Higgsinos can also decay directly to a boson and the gravitino: 
$\widetilde{H}^{0}_{1,2} \to h\,\widetilde{G}$,
$\,\widetilde{H}^{0}_{1,2} \to Z\,\widetilde{G}$,
$\,\widetilde{H}^{\pm} \to W^\pm\,\widetilde{G}$,
 with widths that crucially depend on $\langle F \rangle$.
For the two neutral Higgsino states, these are~\cite{Ambrosanio:1996jn} 
 \begin{equation}
  \Gamma_{i,x}\  \equiv \   
  \Gamma (\widetilde{H}^{0}_{i}\! \to x\,\widetilde{G})
 \ = \ \frac{\vert C_{i,x}\vert^2}{64 \pi}\, 
  \frac{m_{\widetilde{H}^{0}_{i}}^5}{\langle F\rangle^2}
 \Bigg[\!1-\frac{m_x^2}{m_{\widetilde{H}^{0}_i}^2}\!\Bigg]^4\!,
\label{eq:neutwidth}  
\end{equation}
 where it is $i=1,2$; $x=h,Z$; and where, without loss of generality, $\widetilde{G}$
 is assumed to be massless.
In these expressions the coefficients $C_{i,x}$ contain the rotation factors
 from current to mass eigenstates and include the dependences on $\tan \beta$
 and $\sgn\mu$
 ($C_{1,h} = N_{13}\cos\beta  + N_{14}\sin\beta
          = N_{14}[\sin\beta-\sgn{\mu}\cos\beta]$,
 and
  $C_{1,Z} =   N_{13}\cos\beta  - N_{14}\sin\beta
          = - N_{14}[\sin\beta+\sgn{\mu}\cos\beta]$,
 where, for almost-pure Higgsinos, it is
 $\vert N_{1,3}\vert,\, \vert N_{1,4}\vert \simeq 1/\sqrt{2}$ ).
These dependences are shown in Fig.~\ref{fig:BRLFT}(a) for the two
 branching ratios
 $B(\widetilde{H}^0_1 \to Z \widetilde{G})$ and
 $B(\widetilde{H}^0_1 \to h \widetilde{G})$. 
Notice that for 
 $m_{\widetilde{H}^{0}_1}\leq m_h$, it is the {\small{SM}} Higgs that decays
 as $h\to \widetilde{H}^{0}_1 \,\widetilde{G}\to Z\widetilde{G}\,\widetilde{G} $.

%

The  $\sgn \mu$ and $\tan\beta$ dependences are weaker for the total decay width,
 less than 10\% for $m_{\widetilde{H}^{0}_1} > 200\,$GeV and  $\tan\beta > 15$, 
 rendering somewhat admissible in the following the further approximation
 $\sin\beta = 1$ for $\sin\beta \simeq 1$ (needed in the class of models considered
 here in order to reproduce the correct value of $m_h$).
This limiting case is worth exploring as $\sin\beta=1$ eliminates the $\sgn \mu$
 dependence, reducing our parameter space to $m_{\widetilde{H}^{0}_1}$ and
 $\langle F\rangle$ only.
In particular, in this limit it is possible to obtain a
 rather simple expression for the \small{NLSP} lifetime: 
\begin{equation}
  c  \tau_{\widetilde{H}^0_1} \   = \  
  (\Gamma_{1,h}+\Gamma_{1,Z})^{-1}
 \ \simeq \ 
  57 \,{\rm m}
\left(\frac{500\,{\rm GeV}}{m_{\widetilde{H}^0_1}}\right)^5 \times
\left(\frac{\sqrt{\langle F \rangle}}{\sqrt{3}\times 10^7\,{\rm GeV}}\right)^4. 
\label{eq:wlifetime}
\end{equation}

The parameter space used for this study, recast in terms of
 $m_{\widetilde{H}^{0}_1}$ and  $\tau_{\widetilde{H}^0_1}$, is shown in the panel (b)
 of Fig.~\ref{fig:BRLFT}, together with the the dependence of
 $\sqrt{\langle F \rangle}$ on $\tau_{\widetilde{H}^0_1}$
 and $m_{\widetilde{H}^0_1}$, as given in Eq.~(\ref{eq:wlifetime}).
In the following we limit the range of $\langle F\rangle$ in such a way to have
 long-lived {\small{NLSP}}s.
Notice that for $m_{\widetilde{H}^{0}_1}\sim 500\,$GeV and
 $\sqrt{\langle F \rangle} \sim \sqrt{3} \times 10^7\,$GeV, as in the model  
 {\bf P1} of Ref.~\cite{Fukuda:2015pra}, Eq.~(\ref{eq:wlifetime}) 
 gives an {\small{NLSP}} with a lifetime of about $190\,$ns.
Similarly heavy {\small{NLSP}}s, with values of $\sqrt{\langle F \rangle}$
 larger than $10^8\,$GeV are so long lived to fly off outside the
 detector.
 Also shown in Fig.~\ref{fig:BRLFT}(b) are the regions excluded by searches of displaced
 dijets (pink-shaded)  and displaced dileptons (cyan-shaded) at the
 {\small{LHC}} Run-I~\cite{Liu:2015bma}.
The three points denoted by stars pass unscathed all constraints and are 
 used as benchmarks of our following analysis.
(More information on these points can be found in the next sections.)

For less extreme values of $\tan \beta$, say for example $\tan \beta= 15$,
 and smaller values of the {\small{NLSP}} mass, the relation in
 Eq.~(\ref{eq:wlifetime}) needs to be corrected.
The resulting deviation in the position of the $\langle F \rangle$ lines in
 Fig.~\ref{fig:BRLFT}(b) would be, however, only at the $\%$ level and not
 visible in this figure.  
Nevertheless, such corrections will have to be implemented for values of 
 $m_{\widetilde{H}^0_1}$ in the lower end of our mass interval and moderate
 values of $\tan \beta$ that still provide the correct value of $m_h$.
We leave such studies to future work.

\begin{figure}
\begin{center}
   \begin{subfigure}[b]{0.43\textwidth}
    \includegraphics[width=\textwidth]{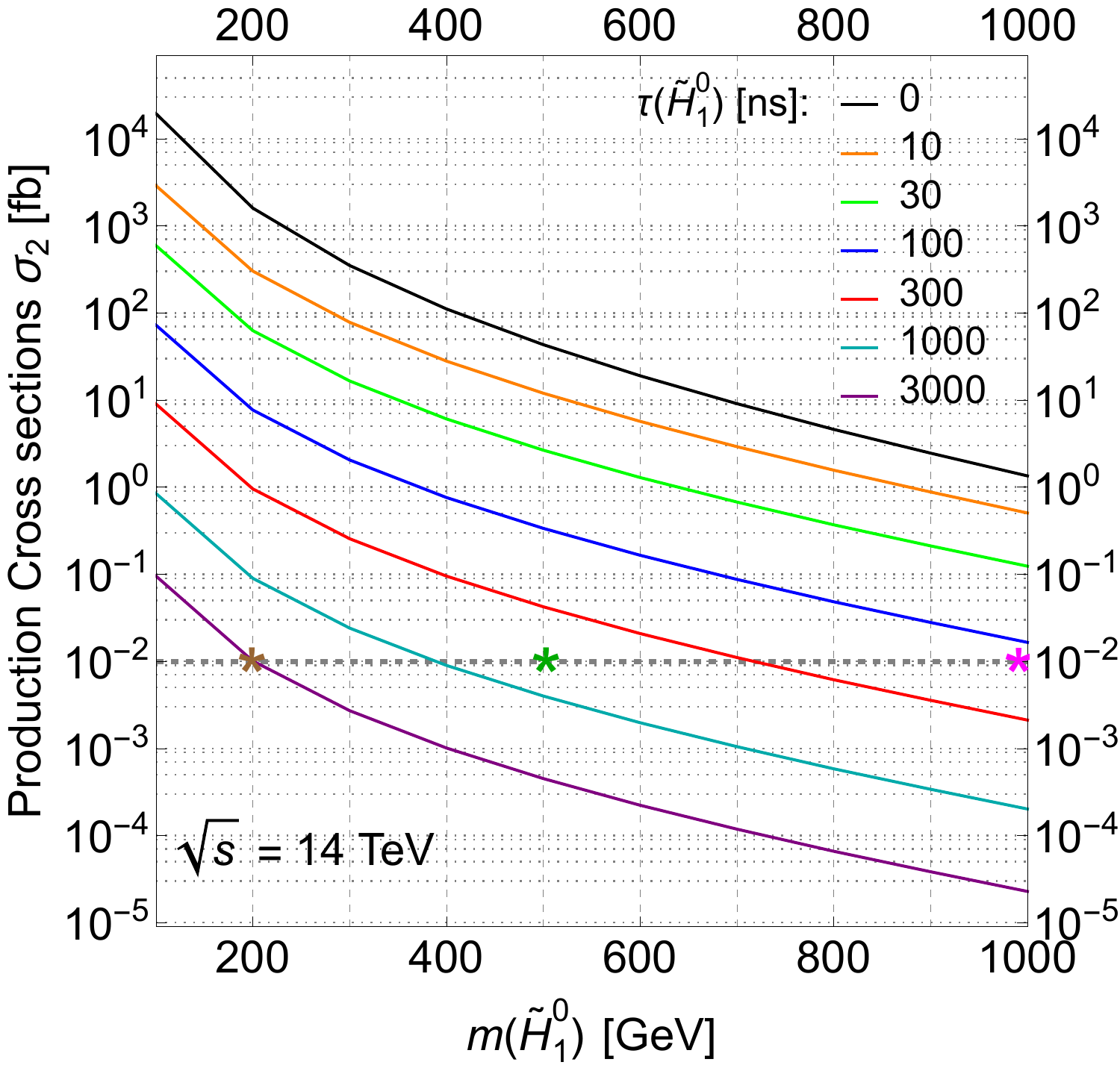}
    \caption{ }
    \label{fig:f2a}
  \end{subfigure}
   \quad  \quad  \quad 
  \begin{subfigure}[b]{0.43\textwidth}
    \includegraphics[width=\textwidth]{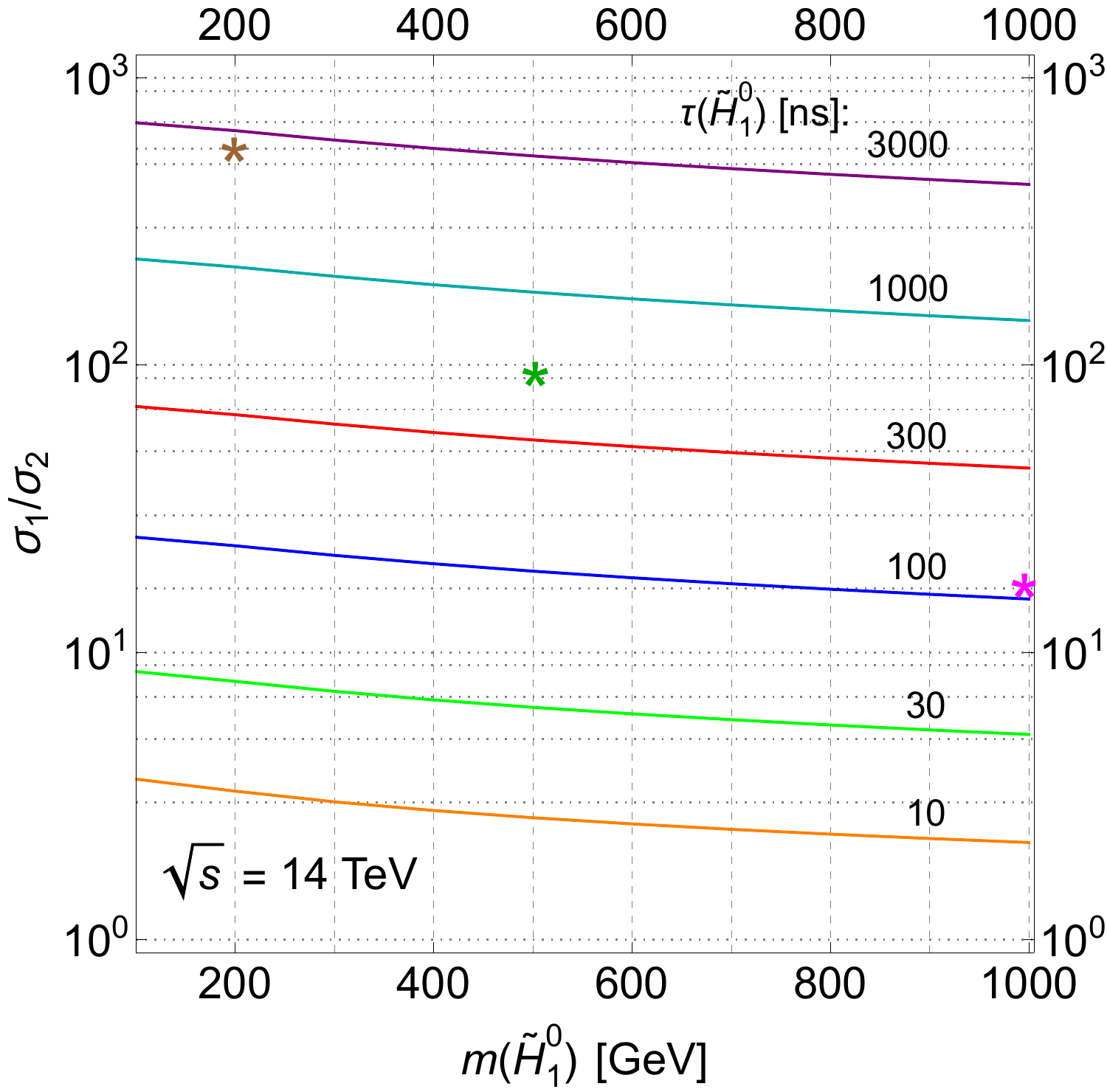}
    \caption{ }
    \label{fig:f2b}
  \end{subfigure}
\end{center}
\caption{\small{Total cross section for the production of a pair of
  {\small{NLSP}}s as a function of $m_{\widetilde{H}^0_1}$ (black curve in 
  panel~(a)).
The cross sections $\sigma_2$ for the case in which both {\small{NLSP}}s decay
  within the fiducial volume are shown in
  the same panel by colored curves for various lifetimes.
The cross sections $\sigma_1$ in which at least one of the pair-produced {\small{NLSP}}s
 decays in this volume can be obtained from the ratios $\sigma_1/\sigma_2$ shown
 in panel (b).
  The three stars correspond to the three benchmark points defined in the text. }} 
\label{fig:XsectRAT}
\end{figure}

{\noindent \bf Production cross section.}
Apart from the superlight gravitino, Higgsinos are the lightest particles
 in the spectrum of \small{SUSY} models with \small{GMSB} and with a
 considerable decoupling of sfermions and gauginos.
Thus, they are necessarily the final products of cascade decays of the
 superpartners produced at sufficiently energetic colliders, such as the
 envisaged {\small{FCC-hh}}~\cite{FCC:2018vvp}
 and the {\small{CEPC-SPPC}}~\cite{Arkani-Hamed:2015vfh}. 
At colliders with a less extreme center of mass energy $\sqrt{s}$, as the
 {\small{LHC}}, the most relevant channels of Higgsino production are
 the Drell--Yan-type of processes, which give rise to:
 $pp \to \widetilde{H}^0_1 \widetilde{H}^0_2$, 
 $\,pp \to \widetilde{H}^{\pm} \widetilde{H}^{\mp}$,  and
 $pp \to  \widetilde{H}^0_i \widetilde{H}^{\pm}$ ($i=1,2)$.
Our typical values of $\Delta m_0$ have here some important consequences:
 {\bf i}) since $\widetilde{H}^0_2$ and $\widetilde{H}^{\pm}$ decay promptly into
 the \small{NLSP}, all the previous processes end up
 into  $pp \to \widetilde{H}^0_1 \widetilde{H}^0_1$;
 {\bf  ii}) the decaying $\widetilde{H}_2^0$ and $\widetilde{H}^{\pm}$
 induce negligible hadronic and leptonic activities, quite difficult to detect
 at the LHC;
 {\bf iii}) from the calculational viewpoint it is possible to assume that 
 all Higgsino states are mass degenerate.
Because also the gauge interactions needed to calculate their production cross sections 
 are equal to those of leptons, the four Majorana spinors composing the Higgsino system
 can then be viewed as the two Dirac spinors of a heavy lepton doublet in a
 vector representation of the SM, and the Higgsino production cross sections
 are equal to those of vector-like heavy leptons.

We have calculated
 $\sigma_0 \equiv \sigma(pp \to \widetilde{H}^0_1 \widetilde{H}^0_1) $
 for the next run of the \small{LHC} with $\sqrt{s}= 14\,$TeV,  
 at the next-to-leading order~({\small{NLO}}) in \small{QCD}, by using 
 MadGraph5\_a\small{MC}@\small{NLO} (\small{MG5aMC})~\cite{Alwall:2014hca}.     
We have implemented an extension of the {\small{SM}} containing one vector-like
 leptonic doublet~\cite{Kumar:2015tna} in FeynRules~\cite{Alloul:2013bka} with the
 \small{NLO} counterterms~\cite{Degrande:2014vpa} needed to create the UFO~(Universal
 FeynRules Output) model file~\cite{Degrande:2011ua} to be plugged in \small{MG5aMC}. 
The resulting cross section is shown as a function
 of the \small{NLSP} mass by the black line in Fig.~\ref{fig:XsectRAT}(a).
For this, as well as for all our results,
 the parton distribution functions \small{NNPDF}3.1~\cite{NNPDF:2017mvq} were used. 
The theoretical uncertainty (due to the renormalization and factorization scales) of all
 cross section shown here is a few percent.

The values obtained for $\sigma_0$ guarantee a copious production of \small{NLSP}
 pairs throughout the whole range of masses considered here.
The identification of these \small{NLSP}s rests on the possibility of
 detecting their decays $Z\widetilde{G}$/$h\widetilde{G}$,  
 within or outside the detector, depending on the values of 
 $m_{\widetilde{H}^{0}_1}$ and $\tau_{\widetilde{H}^0_1}$.
If they decay inside the electromagnetic~(\small{EM}) calorimeter, the total 
 energy of the bosons $Z$ and $h$ may be  measured for most decays.
Timing information of the decay products can be used to efficiently detect
 delayed decays of the \small{NLSP}s~\cite{Liu:2018wte}.
By taking here as example the size of the \small{EM} calorimeter
 in \small{ATLAS}, we set our fiducial volume to be a cylindrical one of
 radius 1.5 m and length 6.4 m.

\begin{wrapfigure}{r}{6.9cm}
  \centerline{\includegraphics[width=0.43\textwidth]{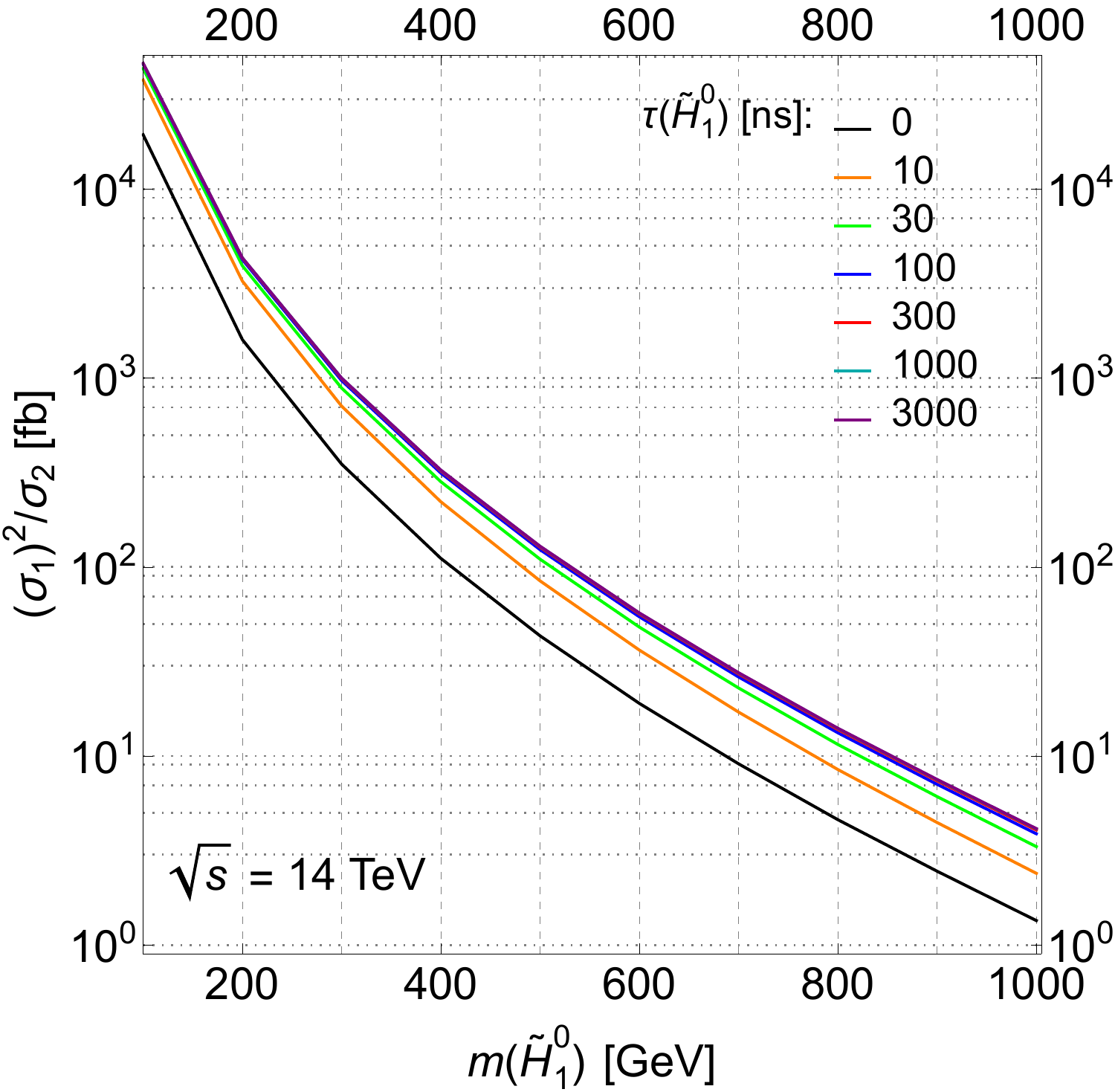}}
  \caption{\small{Mass dependence of the variable $(\sigma_1)^2/\sigma_2$  shown in
      black, orange, green and blue for  $\tau_{\widetilde{H}^0_1}= 0,10,30,100\,$ ns. 
      The curves for  $\tau_{\widetilde{H}^0_1}=300,1000,3000\,$ns are indistinguishable from
      the blue one.}}
\label{fig:MassVar}
\end{wrapfigure}

In the following, we implement three selection criteria for the events of pair
 production, depending on if/where the produced \small{NLSP}s decay.
We select in case 0) all pair-production events;
 in case 1) events in which at least one of the pair of produced \small{NLSP}s
 decays within the fiducial volume; 
 in case 2) events in which both \small{NLSP}s decay within it.
We call  $\sigma_i$ the cross section for the case $i$) ($i=0,1,2$).
We show $\sigma_2$ in Fig.~\ref{fig:XsectRAT}(a) as a function of
 $m_{\widetilde{H}^0_1}$, for different values of $\tau_{\widetilde{H}^0_1}$.
The value $\sigma_2\sim 0.01$ fb, shown in the same figure by 
 the grey-dashed line, corresponds to the
 production of $\simeq 30$ pairs of \small{NLSP}s in our fiducial volume
 with 3000 fb$^{-1}$ data at the high-luminosity \small{LHC}.
The three stars along this line correspond to the three points of parameter space
 chosen later as our benchmark points.
(See next section.)

Note that, for prompt \small{NLSP}s decays 
 (with $\tau_{\widetilde{H}^0_1}\sim 0$), it is $\sigma_2=\sigma_1 = \sigma_0$. 
Moreover, the requirement for case 1)  does not prevent both \small{NLSP}s
 from decaying inside the fiducial volume, as it is in case 2), thus giving 
 values for $\sigma_1$ larger than those for $\sigma_2$ (see  Fig.~\ref{fig:XsectRAT}(b)).
Note also that we cannot directly measure $\sigma_0$, including 
 events where both NLSPs escape the fiducial volume before decay. 

The cross sections $\sigma_1, \sigma_2$ depend on both, $\widetilde{H}^0_1$'s mass and lifetime.  
In contrast, Fig.~2(b) shows that their ratio $\sigma_1/\sigma_2$, obtained just by
 counting the numbers of events for cases 1) and 2), is determined mainly by the
 lifetime, providing therefore a good variable to estimate it.
It appears from Fig.~2(a) that the cross sections $\sigma_2$ decrease for increasing
 lifetime approximately as  $(\tau_{\widetilde{H}^0_1})^{-2}$, whereas the variable
 $\sigma_1/\sigma_2$ in Fig.~2(b) increases roughly as $\tau_{\widetilde{H}^0_1}$.
We deduce that $\sigma_1$ decreases as $(\tau_{\widetilde{H}^0_1})^{-1}$.
These scalings of $\sigma_1$ and $\sigma_2$ with $\tau_{\tilde{H}^0_1}$ 
 are explained as follows: 
 in cases where $\tau_{\tilde{H}^0_1}$ is such that $c\tau_{\tilde{H}^0_1}$
 is sufficiently larger than the size of the fiducial volume, 
 the probability that a produced NLSP decays before escaping the fiducial volume is 
 proportional to $(\tau_{\tilde{H}^0_1})^{-1}$ and, therefore, 
 $\sigma_1\propto(\tau_{\tilde{H}^0_1})^{-1}$ and 
 $\sigma_2\propto(\tau_{\tilde{H}^0_1})^{-2}$.

These observations suggest that the combination $(\sigma_1)^2/\sigma_2$ should be very
 mildly dependent (or altogether independent) from the {\small{NLSP}}
 lifetime, while presumably maintaining a strong dependence on its mass.
An explicit calculation of $(\sigma_1)^2/\sigma_2$ yields the results shown in Fig.~3,
 where  $(\sigma_1)^2/\sigma_2$ is plotted versus $m_{\widetilde{H}^0_1}$.
The colour code for the curves in this figure is that used in Figs.~2(a) and 2(b):
 orange, green, and blue are the curves with value of lifetime equal to 
 10, 30, 100 ns. 
For larger values of lifetime, i.e. 300, 1000, 3000 ns, the various curves for
 $(\sigma_1)^2/\sigma_2$
 collapse into one, undistinguishable from the blue curve.
For  $\tau_{\widetilde{H}^0_1} = 0$,
 $(\sigma_1)^2/\sigma_2$ reduces to $\sigma_0$, shown in this figure, as in
 Fig.~2(a), by a black curve.

Thus, modulo theoretical (rather modest, as mentioned) and experimental errors, 
 a measurement of the variable  $(\sigma_1)^2/\sigma_2$ for long-lived {\small{NLSP}}s
 gives a good estimate of their mass.
Once this is determined, the value of the other variable
 $\sigma_1/\sigma_2$ can help restricting their lifetime.

{\noindent \bf Distributions.}
We analyze here the time, space~($t$, $l$) distributions of the vertices
 of the decaying \small{NLSP}s, as well as the
 boost-factor~($\beta\gamma$) distribution of these \small{NLSP}, all
 normalized:
\begin{equation}
  \frac{1}{n_j}\frac{d n_j}{d t}, \quad \quad  \quad  
  \frac{1}{n_j}\frac{d n_j}{d l}, \quad \quad  \quad  
  \frac{1}{n_j} \frac{d n_j}{d(\beta\gamma)}. 
\label{eq:distrib}
\end{equation}
The various counting functions $n_j$ with $j= 0,1,2$, which define also here the
 cases 0), 1), and 2), are related to the cross sections $\sigma_j$
 shown in  Fig.~\ref{fig:XsectRAT}.
Indeed, $n_0$  counts the number of vertices in which both pair-produced
 \small{NLSP}s decay, irrespective of when and where; $n_2$, counts the
 number of decaying \small{NLSP}s, pair-produced through $\sigma_2$, when both
 decay within the fiducial volume;  $n_1$ counts the \small{NLSP}s produced
 through $\sigma_1$ that also decay within the fiducial volume, and all of
 them, {\it i.e.} one or two for each production event.
In the above expressions, $t$ is the time
 of flight of the decaying \small{NLSP}s in the cases 0) and 1), or the time
 difference between the decay vertices of the pair-produced \small{NLSP}s in
 case 2). 
In the space distributions, $l$ denotes the displacement of the
 \small{NLSP} decay vertex from the primary collision point for cases 0) and 1), or the
 difference of the displacements of the two \small{NLSP} decays for case 2). 
In the $\beta\gamma$ distribution,
 it is  $\beta\gamma=\beta/(1-\beta^2)^{1/2}\!$, $\,\beta = v/c$, and $v$ the
 \small{NLSP} velocity.

\begin{figure}[th!]
\begin{center}
   \begin{subfigure}[b]{0.317\textwidth}
    \includegraphics[width=\textwidth]{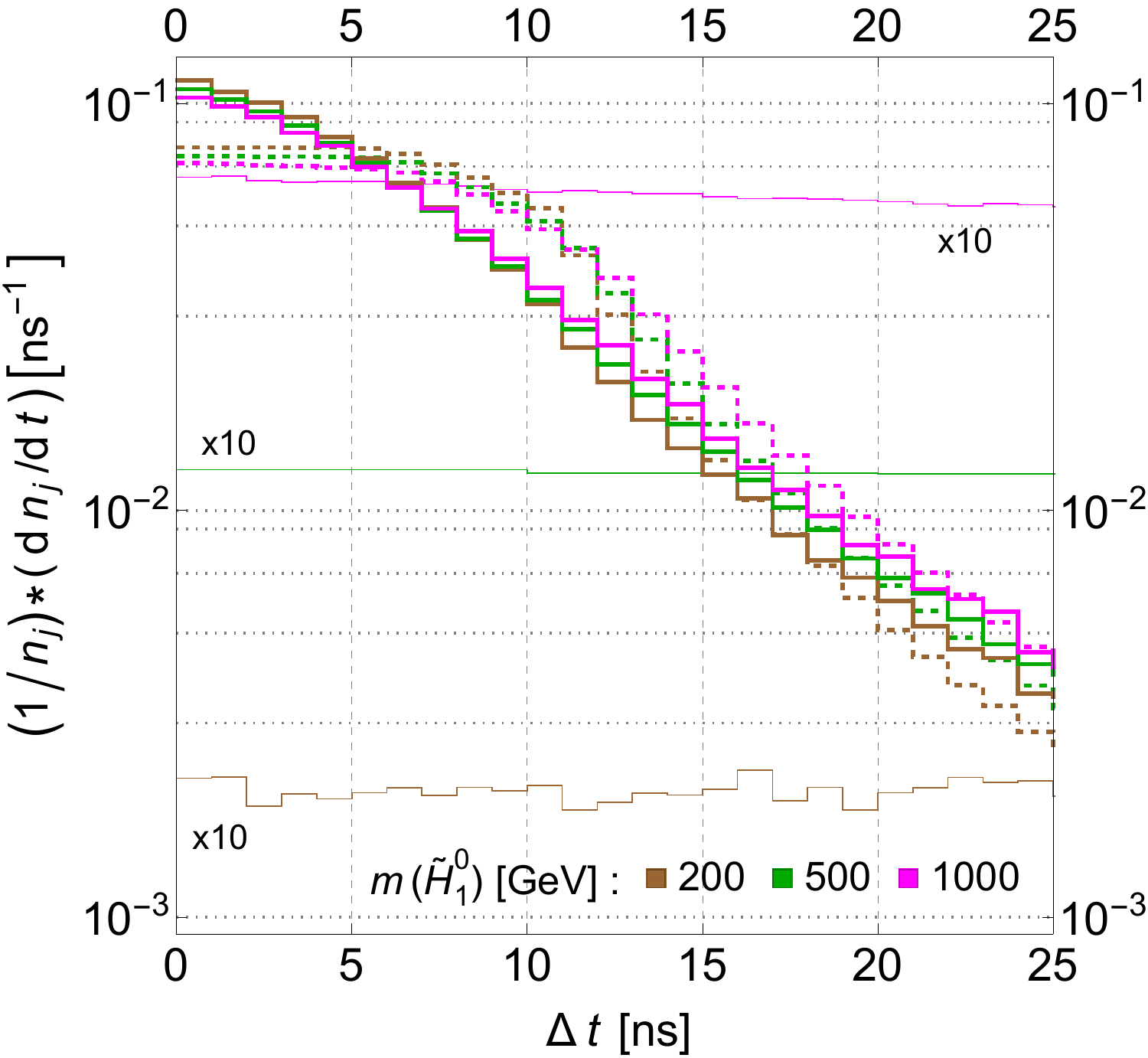}
    \caption{ }
    \label{fig:f3a}
   \end{subfigure}
\quad   
\begin{subfigure}[b]{0.307\textwidth}
    \includegraphics[width=\textwidth]{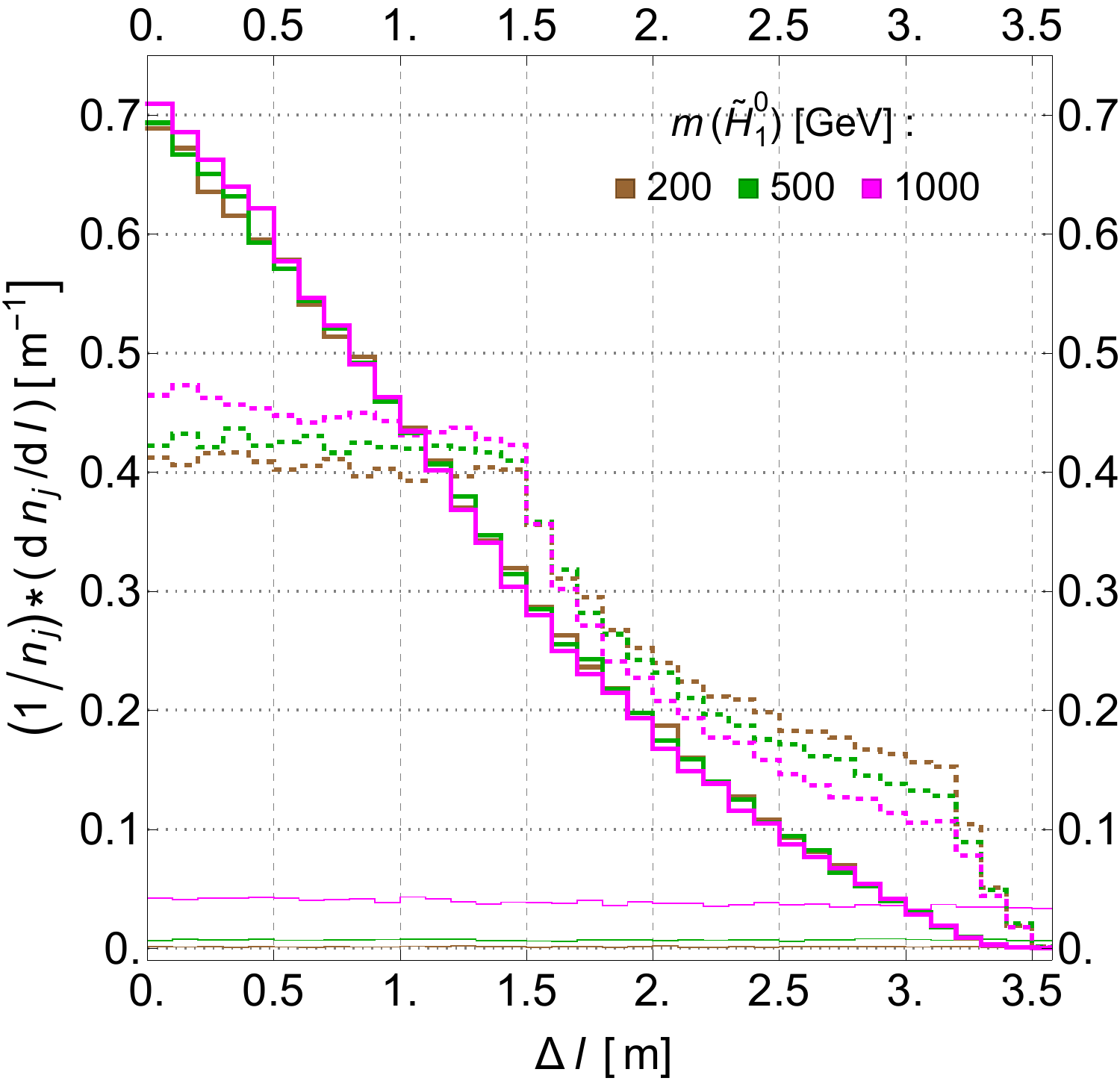}
    \caption{ }
    \label{fig:f3b}
\end{subfigure}
\\[1.0ex]
\begin{subfigure}[b]{0.307\textwidth}
    \includegraphics[width=\textwidth]{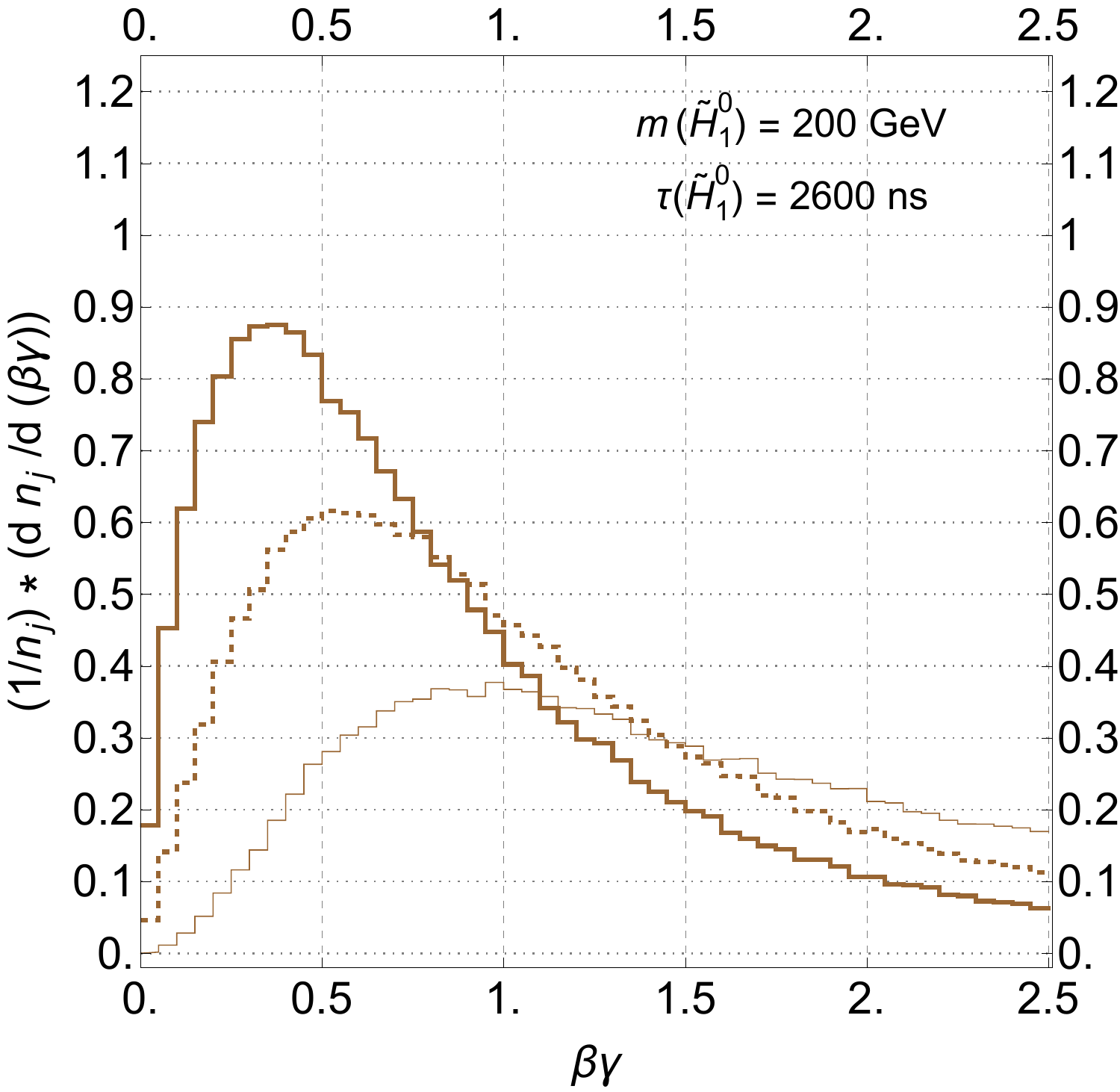}
    \caption{ }
    \label{fig:f3c200}
\end{subfigure}
\quad
\begin{subfigure}[b]{0.307\textwidth}
    \includegraphics[width=\textwidth]{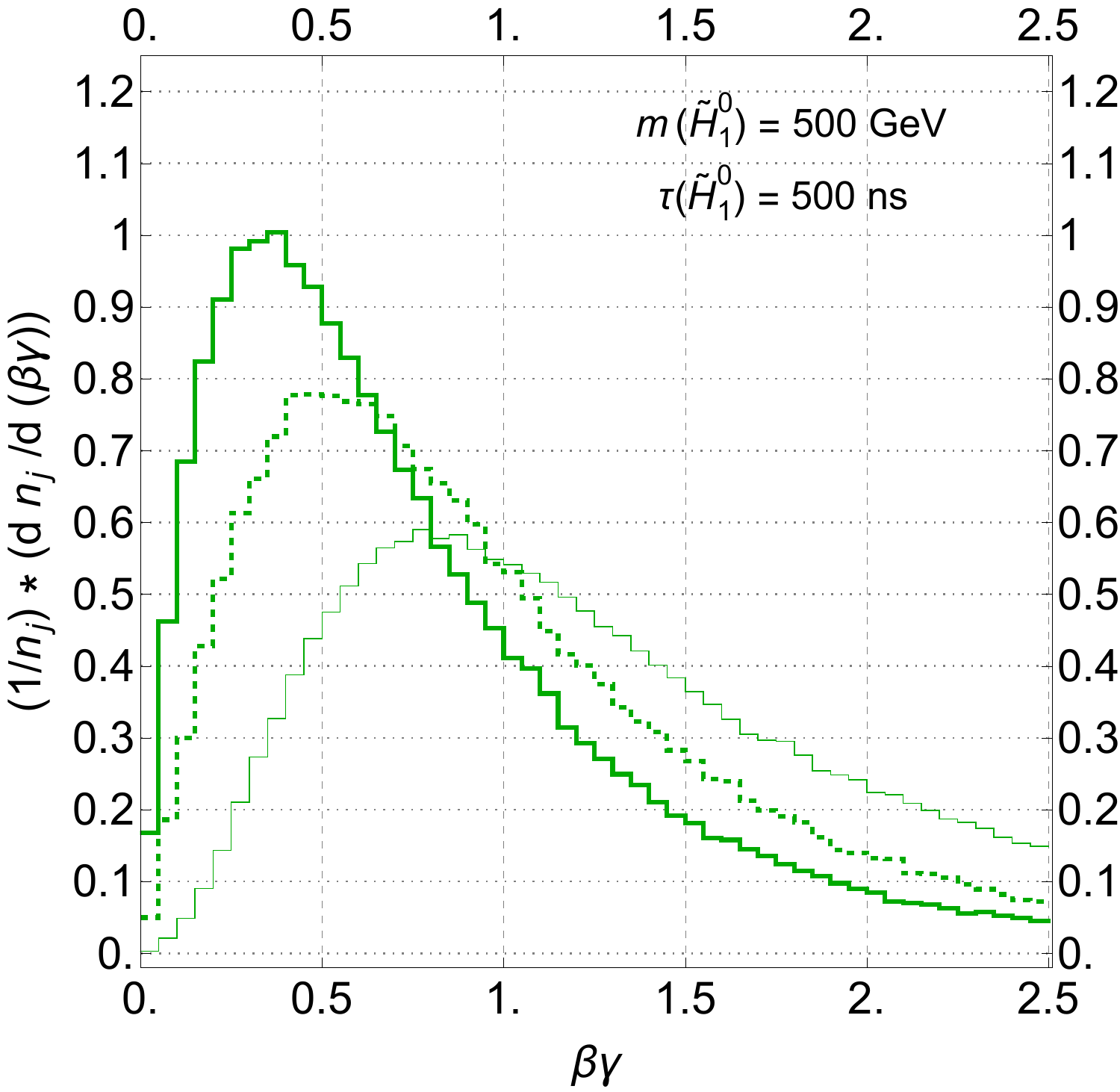}
    \caption{ }
    \label{fig:f3c500}
\end{subfigure}
\quad
\begin{subfigure}[b]{0.307\textwidth}
    \includegraphics[width=\textwidth]{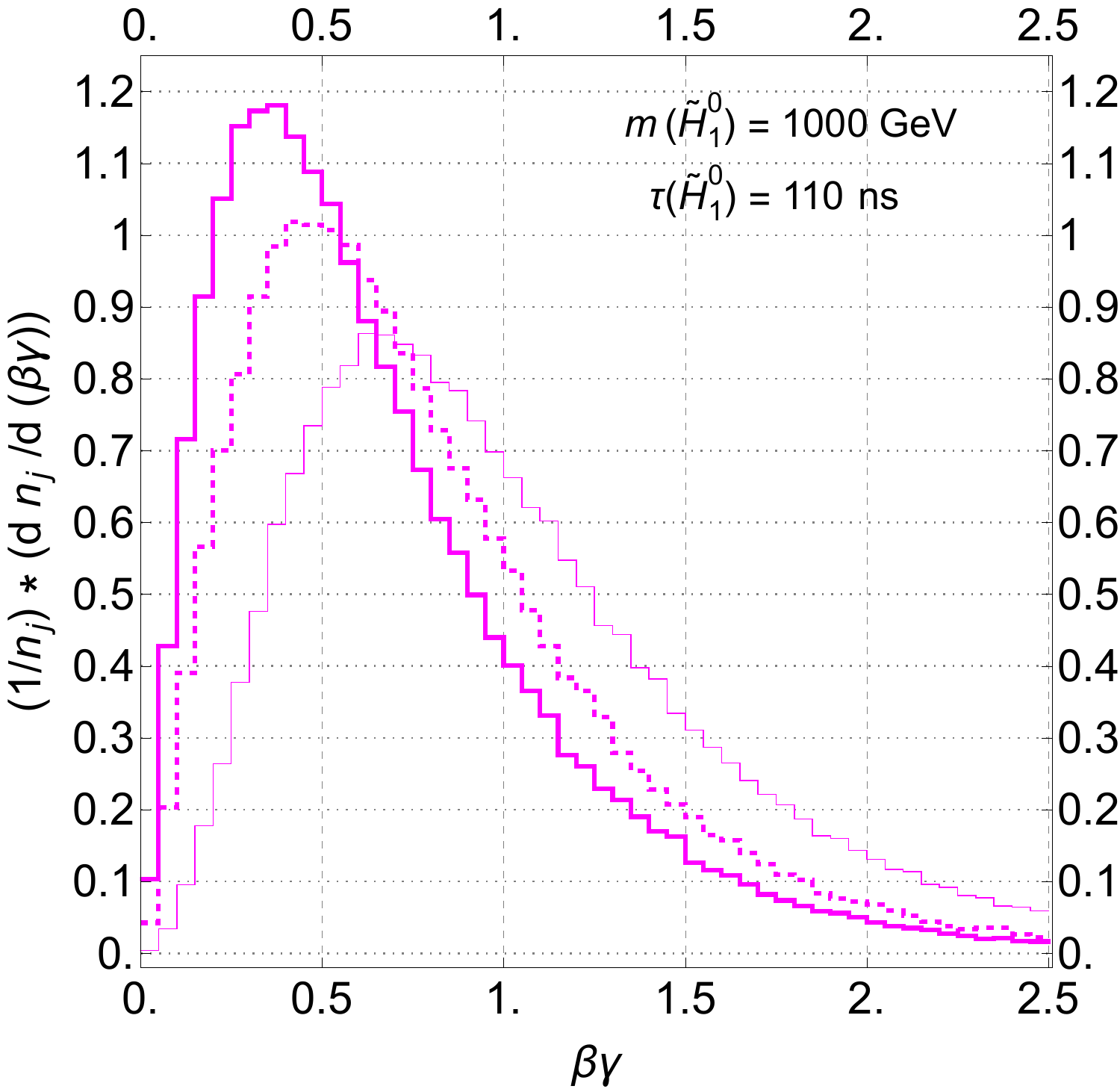}
    \caption{ }
    \label{fig:f3c1000}
\end{subfigure}
\end{center}
\caption{\small{Time (panel (a)), space (panel (b)) and boost-factor (panels (c), (d), (e))
  dependences of the normalized distributions in
  Eq.~(\ref{eq:distrib}) for the {\small{NLSP}}s
  counting cases 0), 1), 2)~(respectively thin-solid, dashed, and thick-solid lines)
  and for three sets of values (mass, lifetime).
In panel (a) the values of the thin lines are 10 times those of
  the actual time distributions, for visibility.}}
\label{fig:XsectDEP}
\end{figure}

 We analyze these distributions for the three benchmark points on
 the grey-dashed line in Fig.~\ref{fig:XsectRAT}(a), corresponding to the values of
 ($m_{\widetilde{H}^0_1}$, $\tau_{\widetilde{H}^0_1}$):
 (200 GeV, 2600 ns), (500 GeV, 500 ns), (1000 GeV, 110 ns).
We show our results in Fig.~\ref{fig:XsectDEP}, where, throughout all panels, we denote
 the distributions by thin-solid/dashed/thick-solid lines for the cases 0)/1)/2).
Panel~(a) collects the time distributions,  panel~(b) the space distributions
 for all three benchmark points, while 
 panels~(c),~(d),~(e) show the boost-factor distributions for each benchmark point.

The distributions corresponding to the case 0) are flat curves in the panels (a)
 and (b) and have values much smaller than those for the cases 1) and 2).
In the same panels, the distributions for the cases 1) and 2) appear very similar
 for the three benchmark points, that is, they are quite insensitive
 to the choice of values for ($m_{\widetilde{H}^0_1}$, $\tau_{\widetilde{H}^0_1}$).
This is a consequence of the size of our fiducial volume, chosen to have radius/length
  $\ll c\tau_{\widetilde{H}^0_1}$.  
In the space distributions in panel~(b), there is sensitivity
 to the size of the fiducial volume in case 1), but not in case 2).

As for the $\beta \gamma$ distributions, shown in panels~(c),~(d),~(e), we obtain similar
 results for the three benchmark points.
It is easy to show that for large enough lifetimes (condition realized in all these
 three points), the normalized distributions depends on $m_{\widetilde{H}^0_1}$, but
 have a negligible dependence on $\tau_{\widetilde{H}^0_1}$.
Thus, the benchmark points, with $m_{\widetilde{H}^0_1}=200,\,500,\,1000\,$GeV, 
 chosen to guarantee a copious production of \small{NLSP} pairs,
 are quite representative of our whole parameter space. 
The distributions for the cases 1) and 2) show softer boost factors 
 compared to those relative to the case 0). 
This is because the selection criteria for the cases 1) and 2), i.e.
 the requirement of decays inside the fiducial volume,
 removes very relativistic \small{NLSP}s, which tend to escape away.
Moreover, the position of the peaks for the cases 1) and 2), determined
 by the slow \small{NLSP}s, is barely dependent on $m_{\widetilde{H}^0_1}$ 
 (in agreement with the behavior of 
 $\sigma_1/\sigma_2$ in Fig.~\ref{fig:XsectRAT}).

We conclude by observing that 
 when we require that both pair-produced \small{NLSP}s decay inside the fiducial volume,
 the initial state radiations tend to be soft.
To show this more explicitly, we list in Table~\ref{tab:1} the ratios, 
 calculated with the help of Pythia8~\cite{Bierlich:2022pfr},
\begin{align}
  R_i=\frac{n_i\ {\rm (with\ }\ H_T> H_{T {\rm min}})}
           {n_i\ {\rm (without\ any\  cut\  on}\ H_{T})}, 
\label{ratio}
\end{align}
 for the three cases ($i=0,1,2$). 
Here $H_T$ is the scalar sum of the transverse energy of all
 hadronic activities,
 coming from the initial-state radiation in our scenario. 
Given the mild mass dependence, we show our results for an \small{NLSP} of
 $200\,$GeV only, for all three cases of event selection, and for three
 different values of the cut $H_{T\rm min}$.
We find that in all cases, the numbers of event rates are higher the lower
 the threshold cuts are.
Of the three event rates, $n_2$ is most sensitive to the cuts, with $n_1$
 being also more sensitive than $n_0$. 
Therefore, in order to observe two decay-in-flight events inside the 
 fiducial volume, an appropriate trigger system is necessary.

\begin{table}[t] 
\center
\begin{tabular}{l|ccc}
\hline 
 & case 0) & case 1) & case 2) \\ 
\hline
 $R_j\,(H_{T {\rm min}}=50$\,GeV) & 0.64 & 0.62 & 0.57 \\
 $R_j\,(H_{T {\rm min}}=100$~GeV) & 0.44 & 0.42 & 0.36 \\
 $R_j\,(H_{T {\rm min}}=200$~GeV) & 0.23 & 0.20 & 0.15 \\
\hline
\end{tabular}
\caption{Ratio of $n_j$ ($j=$0,1,2) with and without a cut $H_{T {\rm min}}$ 
      for $m_{\widetilde{H}^0_1}=200$~GeV.}
 \label{tab:1}      
\end{table}

{\noindent \bf Summary.}
Experimental data seem to suggest that the possibility of finding \small{SUSY} (at
 least a \small{GMSB} \small{MSSM}) may be closely linked to the discovery of the
 long-lived, lightest neutral state $\widetilde{H}^0_1$ of its Higgsino sector,
 the \small{NLSP} of the underlying \small{SUSY} model.
The detection of $\widetilde{H}^0_1$ can provide information on the whole Higgsino
 sector by uncovering the value of its parameters, thus giving some hints for the
 identification of the underlying \small{SUSY} model.

Aiming at finding appropriate variables to help our search, we have studied very
 carefully the cross section for the production of $\widetilde{H}^0_1$ through a
 Drell--Yang type of mechanism.
After some delay, the two \small{NLSP}s produced at the interaction point with
 cross section $\sigma_0$, decay into $h$ or $Z$ and the gravitino $\widetilde{G}$,
 and can deposit large energies within the \small{EM} calorimeter, if they decay
 inside it.
We have considered the cross section in which both \small{NLSP}s, (or at least one of
 them), $\sigma_2$ (or $\sigma_1$) decay(s) in flight within the calorimeter.
We have calculated all these cross sections at the \small{NLO} level in \small{QCD}
 and studied their dependences on the mass and the lifetime.

Trying to maximize the information to be obtained from experimental data, we have
 singled out two variables expressed in terms of the above cross sections: the 
 combination $(\sigma_1)^2/\sigma_2$ and the ratio $\sigma_1/\sigma_2$.
We have found that the first combination is nearly independent on
 $\tau_{\widetilde{H}^0_1}$, while it retains a strong dependence on the mass, whereas
 the ratio $\sigma_1/\sigma_2$, with a rather weak dependence on the mass, maintains a
 dependence on the lifetime.
Thus, the measurement of $(\sigma_1)^2/\sigma_2$ can disclose to a certain accuracy
 (within experimental and theoretical errors) the value of $m_{\widetilde{H}^0_1}$,
 whereas the value of $\sigma_1/\sigma_2$ can help shed some light on
 $\tau_{\widetilde{H}^0_1}$.

The values of the \small{NLSP} mass and lifetime give us  the
 parameters $\mu$ as well as the {\it v.e.v.} $\langle F \rangle$, through  
 Eq.~(\ref{eq:wlifetime}), valid when 
 the approximation  $\sin \beta =1$ for a large value of $\tan \beta$ is appropriate.
When this is not, a measurement of the branching ratios
 $B(\widetilde{H}^0_1\to h \widetilde{G})$ and $B(\widetilde{H}^0_1 \to Z \widetilde{G})$
 can help us in this tradeoff
 by providing some information on the values of $\tan \beta$ and $\sgn \mu$.

We have also investigated the space and time 
 distributions of the vertices of \small{NLSP}s decaying into gravitinos,
 as well as the distribution of the boost factor of the \small{NLSP}s, 
 for three benchmark points of ($m_{\widetilde{H}^0_1}$, $\tau_{\widetilde{H}^0_1}$).
We have done this for the three cases 0), 1), 2) of event selections (irrespective of
 decays, with at least one of the two \small{NLSP}s decaying, or with both of them
 decaying). 
The space and time distributions are quite insensitive to the choice of the
 benchmark point, in spite of their large differences in
 $m_{\widetilde{H}^0_1}$ and $\tau_{\widetilde{H}^0_1}$.
As for the boost-factor distributions, we have shown that, for sufficiently long
 lifetime, the peak position is insensitive to $m_{\widetilde{H}^0_1}$, which is consistent
 with the mild dependence of $\sigma_1/\sigma_2$ on $m_{\widetilde{H}^0_1}$.
For the cases 1) and 2) these distributions are softer than that for the case 0).
This is caused by the fact that the selections for 1) and 2)
 remove too relativistic \small{NLSP}s.

Finally, we have shown that the initial state radiations tend to be softer 
as the event selection gets tighter, from case 0) to 1) and to 2). 
Thus, it is necessary to provide an appropriate trigger system to count the
 events for the cases 1) and 2). 
%

 \vspace{4mm} 
\noindent {{\bf{Acknowledgements}}}\\[1.01ex] 
We thank B.~Allanach, G.~Polesello, Y.~Shimizu, L.~Velasco-Sevilla
 as well as R.~Frederix and B.~Fuks for various inputs. 
The work of K.M. was supported in part by the \small{JSPS KAKENHI} Grants
 No 20H05239, 21H01077 and 21K03583.

\bibliographystyle{JHEP}
\bibliography{BHMYYApr14}

\end{document}